\documentclass[12pt,preprint]{aastex}
\begin{document}

\title{Discovery of the 2010 Eruption and the Pre-Eruption Light Curve for Recurrent Nova U Scorpii}
\author{Bradley E. Schaefer, Ashley Pagnotta, Limin Xiao\affil{Physics and Astronomy, Louisiana State University, Baton Rouge, LA 70803}}
\author{Matthew J. Darnley, Michael F. Bode\affil{Astrophysics Research Institute, Liverpool John Moores University, Birkenhead, CH41 1LD, UK}}
\author{Barbara G. Harris, Shawn Dvorak, John Menke, Michael Linnolt, Matthew Templeton, Arne A. Henden\affil{American Association of Variable Star Observers, 49 Bay State Road, Cambridge MA 02138}}
\author{Grzegorz Pojma\'{n}ski, Bogumi\l ~Pilecki\affil{Warsaw University Observatory, Al. Ujazdowskie 4, 00-478 Warszawa, Poland }}
\author{Dorota M. Szczygiel, \affil{Department of Astronomy, The Ohio State University, 140 W. 18th Ave., Columbus OH 43210}}
\author{Yasunori Watanabe\affil{Variable Star Observers League of Japan, Keiichi Saijo National Science Museum, Ueno-Park, Tokyo Japan}}

\begin{abstract}

We report the discovery by B. G. Harris and S. Dvorak on JD 2455224.9385 (2010 Jan 28.4385 UT) of the predicted eruption of the recurrent nova U Scorpii (U Sco).  We also report on 815 magnitudes (and 16 useful limits) on the pre-eruption light curve in the UBVRI and Sloan r' and i' bands from 2000.4 up to 9 hours before the peak of the January 2010 eruption.  We found no significant long-term variations, though we did find frequent fast variations (flickering) with amplitudes up to 0.4 mag.  We show that U Sco did not have any rises or dips with amplitude greater than 0.2 mag on timescales from one day to one year before the eruption.  We find that the peak of this eruption occurred at JD 2455224.69$\pm$0.07 and the start of the rise was at JD 2455224.32$\pm$0.12.  From our analysis of the average B-band flux between eruptions, we find that the total mass accreted between eruptions is consistent with being a constant, in agreement with a strong prediction of nova trigger theory.  The date of the next eruption can be anticipated with an accuracy of $\pm$5 months by following the average B-band magnitudes for the next $\sim$10 years, although at this time we can only predict that the next eruption will be in the year 2020$\pm$2.

\end{abstract}
\keywords{novae, cataclysmic variables}

\section{Introduction}

Recurrent novae (RNe) are ordinary novae (binary systems with mass accreting onto a white dwarf until thermonuclear runaway is triggered) for which the recurrence time scale is between a decade and a century, such that more than one eruption has been observed (Payne-Gaposchkin 1964; Bode \& Evans 2008; Evans et al. 2008).  To have the fast recurrence time scale, the novae must have the white dwarf near the Chandrasekhar mass and have a high accretion rate.  These properties, at face value, imply that the white dwarf will soon exceed the Chandrasekhar mass and become a Type Ia supernova, and thus RNe are one of the premier candidates for the progenitor class of these supernovae.  RNe typically have relatively fast eruptions, high ejection velocities, and small eruption amplitudes when compared to ordinary novae.  Only ten RNe are known with certainty in our Milky Way (Schaefer 2010).

U Scorpii (U Sco) previously erupted in March 1999 with a peak at V=7.5 mag (Schaefer 2010).  In quiescence, it has V$\approx$17.6 and has deep {\it total} eclipses taking it down to V=18.9 mag (Schaefer 2010) with an orbital period of 1.23 days (Schaefer 1990; Schaefer \& Ringwald 1995).  U Sco is the fastest of all known novae, fading by three magnitudes from peak in just 2.6 days, while its rise from minimum to peak is 6-12 hours (Schaefer 2010).  No light echo was detected to deep limits after the 1987 eruption (Schaefer 1988).  

U Sco has now had ten known eruptions, in the years 1863, 1906, 1917, 1936, 1945, 1969, 1979, 1987, 1999 (Schaefer 2010), and now 2010 as we report in this paper.  With the discovery of the 1917, 1945, and 1969 eruptions (Schaefer 2001; 2004), it has become apparent that U Sco has outbursts at intervals of $10\pm2$ years since 1900.  The exceptions to this are the two intervals of 19 and 24 years, which are easily interpreted as being double intervals, with eruptions around 1927 and 1957 having been missed.  (U Sco is $3.6\degr$ from the Sun every 28 November, so a significant fraction of its very fast eruptions must be missed.)  With this, it became apparent that the next eruption of U Sco should occur in the year $2009\pm2$.  Schaefer (2005) made a better prediction, on the physical basis that the time between eruptions scales as the inverse of the average mass accretion rate between eruptions (as measured from the B-band flux), with the scaling determined by the inter-eruption light curves from prior eruptions.  The predicted eruption date was $2009.3\pm1.0$.  This is the first time that a specific star has been predicted to have an eruption on a specific date.

With this advance notice, a large international collaboration was formed to provide detailed photometry and spectroscopy in the X-ray, ultraviolet, optical, and infrared bands.  With U Sco going from quiescence to peak to one magnitude below peak in 24 hours, we realized that we must have frequent monitoring of U Sco to get a fast alert of an eruption.  To this end, we mobilized daily and hourly photometry with the SMARTS 1.3-m telescope in Chile, the fully-robotic 2.0-m Liverpool telescope (Steele et al. 2004) in the Canary Islands, and the four ROTSE 0.45-m telescopes in Australia, Texas, Namibia, and Turkey.  In addition, we mobilized a large number of observers through the American Association of Variable Star Observers (AAVSO).  For the seven months each year centered on the opposition of U Sco, we got hourly data.  The headquarters of the AAVSO served as the international clearinghouse for discovery reports and delivery of alerts to the world.  In addition, U Sco was heavily monitored from 2001 to 2009 with long time series photometry, where the main goal was to precisely measure the timing of the eclipses.  The result of all this activity from 2000-2010 is the all-time best pre-eruption light curve for any nova.  This paper presents all the magnitudes and an analysis of this large data set.

\section{The Observations}

Since 1987, one of us (BES) has heavily monitored U Sco, with emphasis on the light curve around the time of the eclipses (Schaefer 1990; 2005; 2010; Schaefer \& Ringwald 1995).  These observations have been made with the McDonald 2.7-m, 2.1-m, and 0.8-m telescopes in Texas as well as with the 1.3-m, 1.0-m, and 0.9-m telescopes on Cerro Tololo in Chile.  The typical integration times were 300 seconds in the B-band and I-band and 100 seconds in the V-band.  Normal processing was carried out, and the photometry was done using the IRAF package PHOT, which performs aperture photometry on the stars in this uncrowded field.  The magnitude of U Sco was determined relative to a selection of nearby comparison stars, for which the primary comparison star, named `COMP' (J2000 16:22:25.6 -17:51:34), has B=16.96, V=15.87, R=15.25, and I=14.59 (Schaefer 2010).  The photon statistics, as calculated by PHOT, are generally smaller than 0.01 mag, but the systematic uncertainties, as represented by the scatter in the measures of standard star magnitudes (Landolt 1992; 2009), are typically 0.015 mag.  The quoted uncertainty is the addition in quadrature of 0.015 mag and the uncertainty from photon statistics.  This data set consists of over 2100 magnitudes, mostly as fast time series photometry centered on times of eclipses.  The specific analysis of the eclipse shapes has already been presented in Schaefer (2010), while a specific analysis of the eclipse times is reserved for a separate paper.  For the study in this paper, the eclipse effects would only hide the other variability, so we have not included any magnitudes with orbital phase between -0.10 and +0.10.   In all, we have 162 magnitudes in the U, B, V, R, and I filters from 2001 to 2006.

Starting in early 2008, we (BES, AP, and LX) began frequent regular monitoring of U Sco with the 1.3-m SMARTS telescope on Cerro Tololo.  This telescope is queue-scheduled, so an operator takes images of U Sco for us several times a week, thus allowing long-term frequent monitoring without requiring us to be at the telescope year-round.  Most of the observations were 300 second exposures in the B-band, but we also made several sets of nearly simultaneous BVRI images.  The procedures and analysis were identical to those described in the previous paragraph.  In all, we have 145 magnitudes from early 2008 until late 2009. 

Beginning in early 2008, one of us (BES) started using the robotic ROTSE telescopes to monitor U Sco once every hour.  The ROTSE telescopes (Akerlof et al. 2003) are four automated 0.45-m f/1.9 telescopes with 1.85$\degr$ fields designed to provide very fast response to satellite triggers on Gamma-Ray Bursts.  The four telescopes are located at Coonabarabran, Australia; Mount Gamsberg, Namibia; Bakirlitepe, Turkey; and McDonald Observatory, Texas.  This wide coverage in longitude gives the potential for complete time coverage.  No filters were used, so the resultant magnitudes are similar to a very broad R-band.  The exposure time was 60 seconds in all cases.  The requested cadence was one exposure every hour from every ROTSE telescope, but problems such as clouds, dawn, daylight, a nearby Full Moon, an altitude lower than 20$\degr$, higher priority alerts for Gamma-Ray Bursts, and the usual equipment problems all make for a substantially lower cadence.  In the months around opposition, the ROTSE system achieved the ideal of nearly hourly coverage for around a quarter of the days, while the average coverage was roughly 15 images in every 24 hour interval.  In the months approaching the conjunction of U Sco with the Sun, the daily coverage decreased to one or two images per 24 hour interval.  For example, in 2009, ROTSE first recorded that U Sco was not in eruption on 9 January (43 days after conjunction) and last imaged U Sco on 18 October (40 days before conjunction).  The limiting magnitude varied widely (with clouds, altitude, focus, and the Moon), yet U Sco was visible at low significance on about half the images.  Even on the best images, U Sco did not get better than a 5-sigma detection, so in no case do we have accurate photometry from ROTSE.  In all, we have a set of roughly 7000 useable U Sco images from ROTSE.  One of the goals of the hourly monitoring by ROTSE was so that we (BES, AP, and MT) could frequently check the images to try to discover the eruption as soon as possible.  Another goal was to catch any pre-eruption rise (see Section 4) even if the amplitude was small and the duration was short.  For the eruption in January 2010, there was no significant pre-eruption rise and our (BGH, SD, JM, ML) small-telescope monitoring produced a better light curve than ROTSE.  A third reason for the ROTSE program was the hope that we would catch U Sco on the rise.  In all previous eruptions, U Sco has been recorded on the rise only three times, each being close to the peak, with the rise from quiescence apparently lasting 6-12 hours (Schaefer 2010).  In the hours before the discovery of the 2010 eruption, the Namibia ROTSE telescope did not look at U Sco due to a higher priority follow-up to a Gamma-Ray Burst, the Turkey ROTSE had clouds, and the Australia ROTSE was down with equipment problems.  With the chance lack of any data on the rise and the poor photometric accuracy of the two years of monitoring, we are not presenting any of the ROTSE magnitudes in this paper.

Beginning in early 2008, we (MT and AAH) organized a steady watch on U Sco by the many observers of the AAVSO.  The primary goal was to catch U Sco's eruption as quickly as possible.  The widespread distribution in longitude of the many AAVSO observers makes for frequent monitoring, and this was the best chance of catching the eruption early.  For the half year around opposition, U Sco was checked for outburst up to 6.7 times per day for monthly averages.  A further requirement for getting fast reactions from the world's telescopes was that the discovery had to be communicated from the discoverer to the rest of the world.  For this vital need, the AAVSO Headquarters served as an around-the-clock, every-day-of-the-year communication center.  Observers were instructed to report their discovery electronically, then automated services would alert key individuals who would test for validity and solicit fast confirmation.  Once the eruption was discovered, we would immediately start notifying the world through IAU Circulars and long-prepared phone and email lists.

As part of this effort, many AAVSO members made positive measures of the brightness of U Sco during the pre-eruption phase.  The AAVSO database contains 412 magnitudes (from 29 observers) and 2853 limits (from 102 observers) between the end of the 1999 eruption and the start of the 2010 eruption (JD 2451557.148 to 2455224.127).  The limits were vital at the time of the observation for knowing that U Sco had not erupted, but they are not now helpful for following the accretion rate.  A further 77 magnitudes are not used here, primarily because the photometric system is not standard and the meaning of the magnitude would be unclear.  This leaves us with 335 positive detections in the pre-eruption time interval.  Just over 90\% of these magnitudes were made with unfiltered CCD imaging, where the magnitudes were calibrated differentially from nearby comparison stars using either the V-band or R-band magnitudes.  These magnitudes (designated CV or CR) will not be exactly on either the V or R magnitude systems, but the expected deviations (less than 0.1 mag) are always small compared to normal variations of U Sco.  Our instrumentation is a 16-inch f/10 Schmidt-Cassegrain with a V filter located in New Smyrna Beach, Florida (BGH), an 18-inch Newtonian telescope without filter located in Barnesville, Maryland (JM), and a 10-inch Schmidt-Cassegrain telescope with a V filter located in Clermont, Florida (SD).  During the critical month before the eruption, all positive detections in the AAVSO database are provided by us (BGH and JM) from CCD images.  

Beginning in early 2009, we (MJD and MFB) started monitoring U Sco with the robotic 2.0-m Liverpool Telescope (Steele et al. 2004) at the Observatorio del Roque de Los Muchachos on La Palma in the Canary Islands.  The goals were to define the pre-eruption light curve in many bandpasses and perhaps to catch a pre-eruption rise or the eruption rise itself.  The photometry was all differential with respect to the comparison stars given in Schaefer (2010).  The images were usually taken through many filters in quick succession once each night.  The filters we used were the B, V, Sloan r', and Sloan i'.  Each light curve point consisted of three 60 s exposures.  The data were analysed using Starlink software.  The typical photometric errors had an uncertainty of 0.01-0.02 mag.  In all, we present here 173 magnitudes from the Liverpool Telescope.

An important practical question was whether U Sco erupted during its yearly conjunction with the Sun every 28 November.  The worst case scenario would be for U Sco to go up in early November, fade back to its quiescent level before any detection was made, and for the eruption to be completely missed.  If U Sco went up while behind the Sun, it would be vital to know this so that our community would not be waiting anxiously with many resources, and also so that observations in the late tail could still be performed.  For this, deep images would have to be made as far into twilight as possible.  Professional telescopes do not go low enough in the sky, so the push into twilight was made entirely by AAVSO observers.  For the November 2008 solar conjunction, U Sco was lost on 2 November and deep images showed U Sco to be near quiescence (V$>$16.1) on 3 January, for a solar gap of 62 days.  For the November 2009 solar conjunction, U Sco was last detected at V=18.6 on 21 October (JM), was fainter than 12.0 mag on 4 November (ML), last checked on 6 November, while after conjunction our images showed it to be fainter than V=14.0 on 27 December (ML), fainter than V=14.3 on 28 December (ML), fainter than V=17.4 on 30 December (BGH), and at quiescence (V=17.6) on 4 January (BGH), for a solar gap of 51-59 days.  From the V-band light curve template (Schaefer 2010; Schaefer et al. 2010b), U Sco is at V=16.6 $\sim$42 days after the peak.  With this extremely short duration, the possibility for a missed eruption was for a peak from 3-23 November 2008 or 7-16 November 2009.  As a chance to discover a U Sco eruption in the week around solar conjunction, one of us (SD) used the SOHO LASCO C3 instrument (which hides the Sun behind a white light coronagraph and produces images of stars, comets, and the corona out to 32 solar radii from the Sun) to demonstrate that the nova never came to peak (i.e., V$>$8.6 mag) during that week-long interval.

The last positive detection of U Sco before its eruption was by us (BGH) on JD 2455223.9473.  Nevertheless, there were later useful upper limits.  Observing from the island of Hawaii, we (ML) used a 20-inch f/3.6 reflector to place a visual limit of $>$16.5 mag on JD 2455224.1271.  The ASAS-3N telescope took a V-band image with a 180 second exposure on JD 2455224.1649 and we (GP, DMS, and BP) did not detect any source to a limit of 15.0 mag.  This robotic telescope is located in Maui on Hawaii at an elevation of 3056 m, with a f/2.0 lens with focal length of 200 mm for a field of view of 8.5$\degr$ on a side on a 2048x2048 pixel CCD chip.  The last known observation before discovery (with V$>$9.2) was taken by us (YW) on JD 2455224.3438 with a 60-mm f/5.9 refractor with an unfiltered CCD located in Yokosuka Japan.

From these sources, we have collected 815 magnitudes for U Sco between the end of the 1999 eruption and the discovery of the 2010 eruption.  (We also report on 16 useful limits and the first three eruption magnitudes.)  These are presented in Table 1.  The first column lists the heliocentric Julian Date (HJD) of the observation.  The second column gives the band of the observation.  The third column gives the magnitude and the one-sigma uncertainty.  The fourth column gives the source of the magnitude, either by identifying the telescope or by giving the AAVSO or VSOLJ observer identification code.  HBB is for B. G. Harris, DKS is for S. Dvorak, MJLE is for J. Menke, LMK is for M. Linnolt, SCK is for B. E. Schaefer, and Wny is for Y. Watanabe.  The next column is the orbital phase of U Sco (with primary eclipses at phase 0.0 and 1.0 and secondary eclipses at phase 0.5) for the pre-eruption ephemeris of $HJD=2451234.539+N \times 1.2305470$.  The last column gives the fractional year corresponding to the Julian Date of the observation.  Figure 1 displays the full light curve for the B filter, while Figure 2 shows the V band light curve after the start of 2009.  Figure 3 shows the folded light curves for the B and V bands.

Table 2 lists various characteristic quantities (magnitudes, colors, and fluxes), their averages, their RMS scatters, and the number of observations going into the averages.  For the science of this paper, we are only interested in the non-eclipsing behavior, so to be conservative, we have included only measures more than 0.10 in phase (0.123 in days) away from the central eclipse times (i.e., between phases 0.10 and 0.90).  For the colors, we have included only those colors derived from two magnitudes taken within 0.005 days of each other so as to keep the errors introduced by fast variations to a minimum.  The overall magnitudes (for all bands) and colors for the entire time interval (from 2000 to the 2010 eruption) are recorded.  We also break up the B and V band magnitudes into smaller intervals so as to seek significant variations.  Finally, we include the average B-band fluxes as defined in Section 6.

\section{Discovery of the 2010 Eruption}

The 2010 Eruption was discovered by us (BGH and SD) as part of systematic nightly monitoring aimed specifically at the discovery of the eruption.  Harris imaged U Sco at 2010 Jan 28.4385 UT (JD 2455224.9385), saw the bright star in the center of the field, and quickly realized that U Sco was in eruption.  Her first act was to send the observation to the AAVSO, and then she telephoned Schaefer.  Schaefer could not get confirmation from ROTSE, so he took his 6-inch telescope out into the front yard and made direct visual confirmation that U Sco was bright in eruption.  Independently, Dvorak discovered the eruption, notified the AAVSO, and started a time series on U Sco to cover the short time interval until dawn got too bright to continue.  These initial observations are included in Table 1.  Circumstances, pictures, and anecdotes on the two independent discoveries are given in Simonsen \& MacRobert (2010).

In practice, our organization worked perfectly.  The AAVSO automated alert system woke up MT and AP.  Within an hour of the discovery, the eruption had been confirmed and worldwide notifications were started.  The first was to the IAU Circulars (Schaefer et al. 2010a).  The sun had already risen in Chile, so we started with more western observatories as well as spacecraft.  Within two hours, BES, AP, and MT had worked through all the long-prepared contact lists.  The response to these contacts (both by members of our existing collaboration as well as by independent observers) was excellent and fast.

The discovery of the 2010 eruption was a fulfillment of the prediction in Schaefer (2005) that U Sco would next erupt in the year 2009.3$\pm$1.0.  The eruption in 2010.1 falls well within the one-sigma region of the prediction.  This adds good confidence to the physical method of summing the total accreted material based on the B-band flux in the prior inter-eruption interval.

\section{Variations in the Light Curve}

The folded light curve (see Figure 3) shows the primary eclipse at phases 0.0, 1.0, and 2.0.  (The 
magnitudes are double plotted so as to make the eclipse at phase 1.0 easily visible.)  The out-of-eclipse brightness varies substantially, and this makes for a ragged eclipse light curve because each point is from a different epoch eclipse with a different amount of flickering light added.  The scatter around the middle of the eclipse is much smaller than the out-of-eclipse scatter, which implies that the flickering region is small and centrally located.

No secondary eclipse is visible in the B and V bands.  However, in the I-band, the secondary eclipse is visible with amplitude roughly 0.3 mag.  This is readily understood as the companion star is much cooler than the accretion disk so eclipses of the companion can only become noticeable at longer wavelength.

All cataclysmic variables, including novae and recurrent novae, show fast flickering.  U Sco is no exception, and this flickering causes the substantial scatter in Figures 1 and 2.  To quantify this, we have calculated the magnitude difference between pairs of magnitudes in the same band, with the pairs being separated in time by some range of delays.  When the delays are shorter than one hour, the RMS scatter of the magnitude differences is 0.06 mag, which is consistent with the expected scatter as based only on the quoted error bars.  When the delays are longer than one day, the RMS scatter of the magnitude differences is 0.27, which corresponds to no correlation between the flickers.  The RMS scatters are 0.09, 0.13, 0.18, 0.19, and 0.21 mag for delays of 0.05-0.10, 0.10-0.15, 0.15-0.20, 0.20-0.50, and 0.50-1.00 days, respectively.  With this, the timescale for correlated variations is from one hour to one day.  The amplitude of these variations is given by the maximum values of the magnitude differences, which is roughly 0.4 mag for all delays longer than 0.05 days.  Our observations are not sensitive to fast, small-amplitude variations.

Many of the recurrent novae have large secular variations in their quiescent light curves (Schaefer 2010).  However, U Sco does not appear to have any long-term variations, as can be seen in Figure 1.  To quantify this for the B and V light curves, Table 2 gives the averages for various time intervals.  Again, no significant variations on time scales of one year or longer are found, despite having small error bars due to the many magnitudes included in the averages.  On the timescale of a tenth of a year, there is marginal evidence for variations, with the V-band average for 2009.7-2009.9 being $0.27\pm0.05$ fainter than average, but this variation is not reflected in other bands, so we think that this apparent change is not significant.  In all, we see no evidence for variations on timescales from 0.1 to 10 years.

On longer times scales, U Sco has small, marginally-significant variations.  During the last four inter-eruption intervals, the average B magnitude was 18.44$\pm$0.07, 18.30$\pm$0.05, and 18.52$\pm$0.04, and 18.45$\pm$0.02 for 1969-1979, 1979-1987, 1987-1999 (Schaefer 2005), and 1999-2010 respectively.  The chi-square for the observed averages (on the hypothesis that the average is a constant) is 12.1 for three degrees of freedom.  The best argument for the significance of these variations is that the deviations from the average are correlated with the duration of each inter-eruption time interval as predicted by theory (see Section 7).

The earliest recorded image of U Sco in quiescence is from the original Palomar Sky Survey.  We measure B=18.80$\pm$0.15 and R=18.00$\pm$0.18 for 29 June 1954.  This is substantially fainter than is normal for later decades, and the time is far from any plausible eclipse.

\section{No Pre-eruption Rise or Dip}

Robinson (1975) examined all known pre-eruption light curves of novae as based on reports in the literature.  He found that five out of eleven novae have pre-eruption rises, lasting months to years in advance of the eruption, with amplitudes from 0.15 to 1.5 mag.  Collazzi et al. (2009) have gone to the original archival photographic plates to measure many pre-eruption light curves, including the key novae with claimed pre-eruption rises.  Four of the five claimed pre-eruption rises were found not to exist as based on our examination of the original plates, such that the claimed rises were caused by simple errors in the literature. Nevertheless, one of the rises (for V533 Her) was confirmed and extended, with the rise being an exponential increase over $\sim$1.5 years and a brightening by up to 1.5 mag.  Also, an additional pre-eruption rise was confirmed in V1500 Cyg, which brightened from roughly 21.5 mag to roughly 13 mag in the month preceding the eruption.  (V1500 Cyg had the fastest known {\it classical} nova eruption, and was second only to U Sco itself amongst all novae.)  In addition, a complex pre-eruption dip was confirmed for the recurrent nova T CrB in the year before the eruption, with the dip being 1-2 mag deep.  In all, three out of 22 novae had either pre-eruption rises or pre-eruption dips.

With other recurrent novae and the fastest classical nova having anticipatory rises and dips, we should investigate whether U Sco has any similar changes.  (We were also hopeful that any such phenomenon would allow us to anticipate the next eruption.)  For this, we can use our light curves to seek any rises or dips.  A glance at Figure 1 quickly shows that there is no significant rise or dip.  Quantitatively, Table 2 shows that the B and V magnitudes from 2010.0-2010.1 are not significantly high or low. And looking at the bottom of Table 1, we see that the last positive detection of U Sco has V=18.2 (BGH) just 24 hours before the discovery.  With this, we can put strong limits on the presence of any pre-eruption rise or dip to be less than roughly 0.2 mag in amplitude on time scales from one day up to a year.

\section{The Rise to Peak}

U Sco rises from quiescence to peak in roughly 6-12 hours, although this is based on just three pre-peak  positive detections and one limit (Schaefer 2010).  For the 1936 eruption, a single Harvard plate shows B=10.75 at a time 0.25 days before peak.  For the 1987 eruption, Dr. N. W. Taylor measured V=14.0 in the day before the peak.  For the 1999 eruption, P. Schmeer measured V=9.5 at a time 0.31 days before the peak.  The final rise to maximum is at a rate of around 19 magnitudes per day, which would imply a rise time of roughly 12 hours if the rise is uniform throughout.  For the 1999 eruption, B. Monard set a limit that $V>14.3$ at a time 0.46 days before the peak.

We can use our data to obtain the best estimate for the time of peak in the 2010 eruption.  The discovery image was taken at JD 2455224.9385 (BGH), and we have measured the magnitude of U Sco (with respect to the comparison star sequence in Schaefer 2010) to be V=7.85 with the systematic uncertainties dominating at around 0.10 mag.  The observed initial rate of decline is 1.4 mag per day (Schaefer et al. 2010b).  The peak magnitude of U Sco is V=7.5, primarily as based on the observed peak of the 1999 eruption (Schaefer 2010).  In an exhaustive comparison of all RN eruptions and those from U Sco in particular, Schaefer (2010) found that all RN are consistent with having the identical eruption light curve shapes, and this is our basis for taking the peak from the 1999 U Sco eruption as being the same for the 2010 eruption.  With this, the peak of the 2010 eruption would have been 0.25 days before discovery, which gives a peak at JD 2455224.69 with a likely uncertainty of 0.07 days.  

The observational limits from the 2010 eruption show that the eruption could not have started much before the ASAS-3N image at JD 2455224.1649.  From the limits on the prior eruptions, the eruption started 0.25 to 0.5 days before the peak, which is roughly from JD 2455224.19 to 2455224.44.  Thus, the time of the start of the expansion, as required by the `universal decline law' of Hachisu \& Kato (2006), can be expressed as JD 2455224.32$\pm$0.12.

Disappointingly, we have no observations from JD 2455224.3438 to 2455224.9385 and thus have completely missed the entire rise and the hour of peak.  In January, U Sco is fairly close to the Sun and hence only visible from a narrow slice of longitude at any given time.  In the southern hemisphere, the start of the rise would only have been visible from the longitudes in the Indian Ocean, while the peak would only have been visible from the longitudes in the South Atlantic Ocean.

\section{Predicting the Next Eruption}

Schaefer (2005) presented a new method for predicting the date of the next eruption of a recurrent nova based on the requirement that some constant amount of mass must be accumulated by the white dwarf between eruptions.  Accretion rates vary substantially (Schaefer 2010; and see Figure 1), so the interval between eruptions ($T$) depends on the average accretion during that time.  If the accretion rate is high then the interval will be short, while if the accretion rate is low then $T$ will be high.  For U Sco, the blue light is dominated by the accretion disk, so the blue flux ($F_B$) will be a measure of the accretion rate.  In particular for U Sco, the accretion rate will be proportional to $F_B^{1.5}$ (Schaefer 2005).  By averaging $F_B^{1.5}$ over each interval $T$, we can derive a quantity that is proportional to the average accretion rate.  Then, $\langle F_B^{1.5} \rangle T$ should be proportional to the total mass accreted between eruptions, which should be a constant.  Schaefer (2005) found that this quantity is indeed constant for four intervals for T Pyx and three intervals for U Sco (despite widely varying values of $T$ for each system), with this providing a good test of nova trigger theory.  These {\it observed} values for $\langle F_B^{1.5} \rangle T$ provide empirical measures of the mass required to trigger the eruptions.  Then, based on the observed B magnitudes up until 2005, Schaefer (2005) was able to predict that U Sco would next erupt in $2009.3\pm1.0$.  Schaefer (2010) updated the situation to arrive at the same predicted date.  As noted above, the actual eruption on 2010.1 falls well within this prediction.

Now, with the full pre-eruption light curve, we can better test the prediction and we can refine the constant for use in predicting the next eruption.  To this end, we have first converted the B-band magnitudes to flux units where B=18 is taken to be the unit flux ($F_{B,18}$), which equals $10^{(18-B)/2.5}$.  The accretion rate will then be proportional to $F_{B,18}^{1.5}$.  To get the time averaged value from 2000-2010.1, we should not simply average all the values, as this would produce a high weight to the behavior of U Sco during 2008 and 2009 (during which the majority of the B-band magnitudes were taken).  Instead, we have taken time intervals and combined them with weights given by their duration.  In Table 2, we list the average values of the measured $F_{B,18}^{1.5}$ for three intervals with roughly constant frequency of observations.  The uncertainty in these averages is the RMS scatter divided by the square root of the number of observations.  The durations of these intervals were then used as weights for averaging the intervals, and the uncertainty in the overall average comes from the usual propagation of errors.  The resultant average over the entire interval between the 1999.2 and 2010.1 eruption is listed in Table 2.  In all, $\langle F_{B,18}^{1.5} \rangle = 0.577 \pm 0.045$.

Schaefer (2005) produces values of $\langle F_B^{1.5} \rangle$ for the three prior inter-eruption intervals.  These need to be updated for four reasons.  First, we need to standardize to the same flux level (i.e., B=18) as the unit flux.  Second, we should be consistent and not include any magnitudes within phase 0.10 of the eclipse.  Third, the individual observations should be formed into the averages with equal weight (instead of weighted by the measurement uncertainty) because intrinsic fluctuations are substantially larger than the measurement errors (so we would not want to give high weight to a bright point simply because it has a small error bar).  Fourth, the 1969-1979 interval has only two magnitudes, so instead of determining the uncertainty based on the RMS scatter of just these two, we have equated the scatter to that during the 1987-1999 interval.  The resultant $\langle F_{B,18}^{1.5} \rangle$ values are given in Table 2.

U Sco erupted in the years 1969, 1979, 1987, 1999, and 2010, with $T$ values of 10.4, 7.9, 11.8, and 10.9 years for the four inter-eruption intervals.  The longest interval is a factor of 1.5$\times$ the shortest interval.  We see that the shortest interval has the highest average accretion rate, while the longest interval has the lowest average accretion rate, and the two middle intervals have the middle average accretion rates.  The four intervals in time order have $\langle F_{B,18}^{1.5} \rangle T$ values of $5.7\pm1.2$, $5.2\pm0.4$, $6.0\pm0.4$, and $6.3\pm0.5$.  (The weighted average of these four values is 5.77$\pm$0.24.)  Nova trigger theory predicts that these values should be a constant.  Indeed, the chi-square equals 3.4 for the hypothesis that the four values are equal to a constant, which is acceptable given the three degrees of freedom.  And we see that the values are consistent with being a constant despite $T$ varying by up to a factor of 1.5.  So we have an improved confirmation of nova trigger theory.

U Sco erupted in 1945 and 1969, with an inter-eruption interval of 23.7 years.  The long interval could be because {\it one} eruption was missed around 1957 (with intervals of around 11.8 and 11.9 years) or because {\it two} eruptions were missed around 1953 and 1961 (with intervals or around 7.9, 7.9, and 7.9 years).  These two possibilities can be distinguished due their greatly different prediction as to the quiescent B magnitude, roughly 18.52 versus 18.30 respectively.  The one measured magnitude (B=18.80$\pm$0.15 from 1954.5) suggests that the accretion rate was low, and hence that there was only one missed eruption.

When will the next eruption of U Sco occur?  Over the next decade, we can keep track of the B-band magnitudes and work out when $\langle F_{B,18}^{1.5} \rangle T$ will equal $5.77\pm0.24$.  Such a prediction will only be accurate to roughly 5 months out of ten years, which is fairly good.  However, this method cannot be used yet, because we cannot predict the variations in the U Sco accretion rate.  For now, the best that we can do is to use the long record of U Sco where all of its inter-eruption intervals are $10\pm2$ years.  With this, we predict the next U Sco eruption to be in 2020$\pm$2.

~

We thank the many observers from the American Association of Variable Stars Observers for their data as used in our light curves.  This work is supported under a grant from the National Science Foundation (AST 0708079).  The Liverpool Telescope is operated on the island of La Palma by Liverpool John Moores University in the Spanish Observatorio del Roque de los Muchachos of the Instituto de Astrofisica de Canarias with financial support from the UK Science and Technology Facilities Council.

{}

\begin{deluxetable}{llllll}
\tabletypesize{\scriptsize}
\tablewidth{0pc}
\tablecaption{U Sco Pre-eruption Light Curve}
\tablehead{\colhead{HJD} & \colhead{Band} & \colhead{Magnitude} & \colhead{Source} & \colhead{Phase} & \colhead{Year}}
\startdata
2451702.1932	&	CV	&	18.1	$\pm$	0.1	&	AAVSO (LAQ)	&	0.038	&	2000.432	\\
2451708.1195	&	CV	&	18.6	$\pm$	0.1	&	AAVSO (LAQ)	&	0.854	&	2000.448	\\
2451719.0413	&	CV	&	18.8	$\pm$	0.1	&	AAVSO (LAQ)	&	0.729	&	2000.478	\\
2452016.8446	&	V	&	17.90	$\pm$	0.01	&	McDonald 2.1m	&	0.738	&	2001.293	\\
2452018.8356	&	V	&	18.05	$\pm$	0.02	&	McDonald 2.1m	&	0.356	&	2001.299	\\
2452018.8397	&	R	&	17.69	$\pm$	0.01	&	McDonald 2.1m	&	0.359	&	2001.299	\\
2452021.7832	&	R	&	17.61	$\pm$	0.02	&	McDonald 2.1m	&	0.751	&	2001.307	\\
2452021.7864	&	R	&	17.63	$\pm$	0.01	&	McDonald 2.1m	&	0.754	&	2001.307	\\
\enddata
\label{Table1}
\end{deluxetable}

\begin{deluxetable}{lllll}
\tabletypesize{\scriptsize}
\tablewidth{0pc}
\tablecaption{Average Magnitudes, Colors, and Fluxes}
\tablehead{\colhead{Quantity}	& \colhead{Date Range}	&	\colhead{Average}	&	\colhead{RMS}	&	\colhead{Number}	}
\startdata
U	&	2004.5	&	18.13	&	\ldots	&	1	\\
B	&	2001.6-2010.1	&	18.45	&	0.26	&	145	\\
V	&	2000.4-2010.1	&	18.01	&	0.29	&	229	\\
R	&	2001.2-2009.8	&	17.67	&	0.26	&	127	\\
I	&	2001.5-2009.7	&	17.35	&	0.14	&	115	\\
Sloan r'	&	2009.3-2010.1	&	17.86	&	0.18	&	26	\\
Sloan i'	&	2009.1-2010.1	&	17.53	&	0.20	&	33	\\
U-B	&	2004.5	&	-0.27	&	\ldots	&	1	\\
\enddata
\tablenotetext{a}{The quoted value is not the RMS scatter of the quantity, but rather is the one-sigma uncertainty in the average value.}
\label{Averages}
\end{deluxetable}

\clearpage
\begin{figure}
\epsscale{1.0}
\plotone{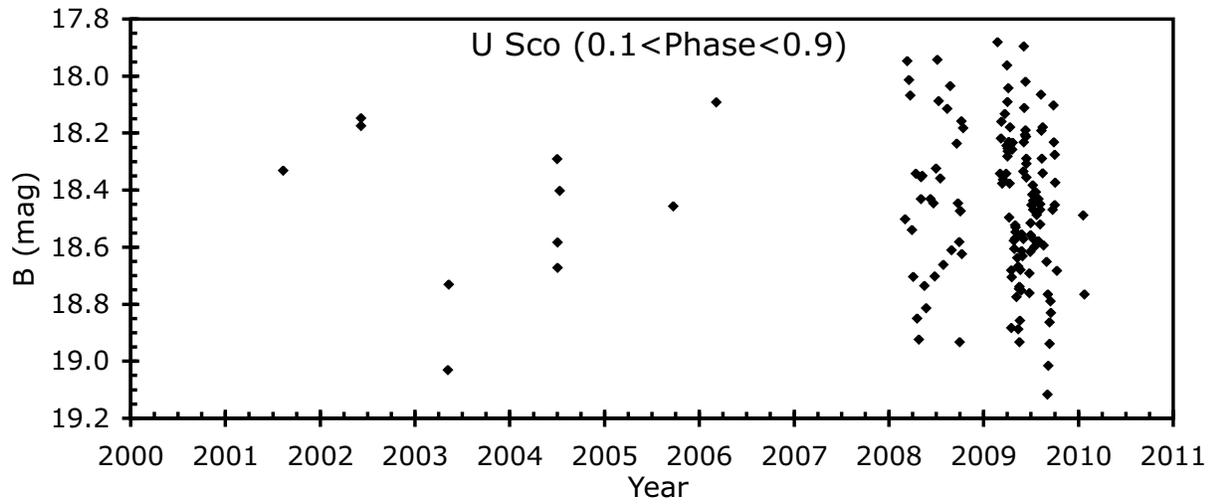}
\caption{
U Sco light curve in B, from 2000 to 2010.  The B-band light comes almost entirely from the disk and is a measure of the mass accretion rate.  This light curve does not include observations within 0.1 phase of the eclipses.  The light curve shows substantial short-term changes but no significant long-term variations.}
\end{figure}

\clearpage
\begin{figure}
\epsscale{1.0}
\plotone{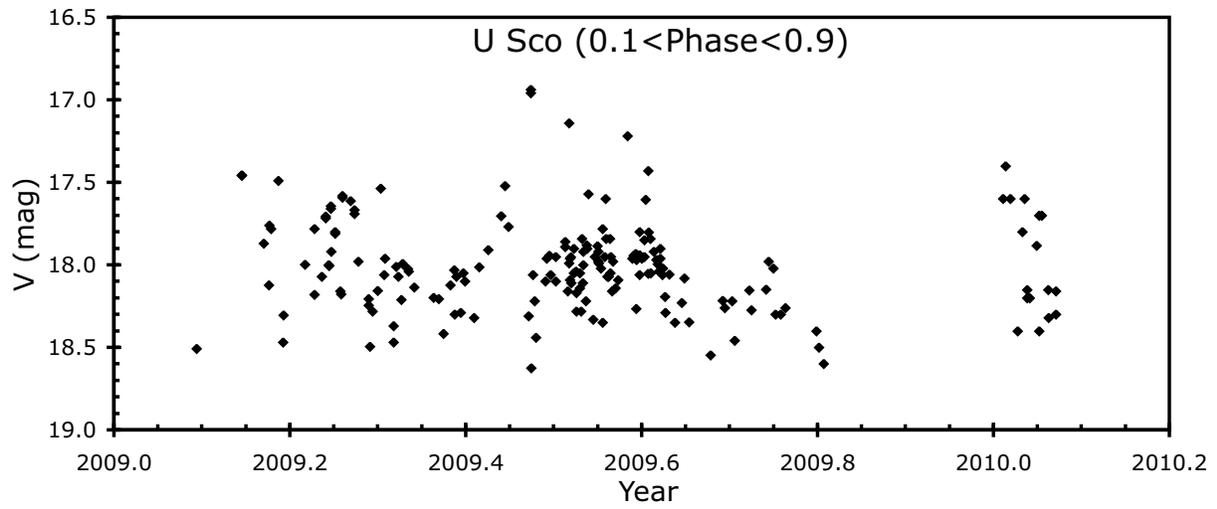}
\caption{
U Sco light curve in V for the last year before eruption.  The observations within 0.1 phase of eclipses are not included so as to concentrate on changes of the system brightness alone.  U Sco shows frequent short timescale variations, but long-term changes are apparently not significant.  In particular, U Sco does not show any pre-eruption rise or dip on timescales from one day to years before the eruption.}
\end{figure}

\clearpage
\begin{figure}
\epsscale{1.0}
\plotone{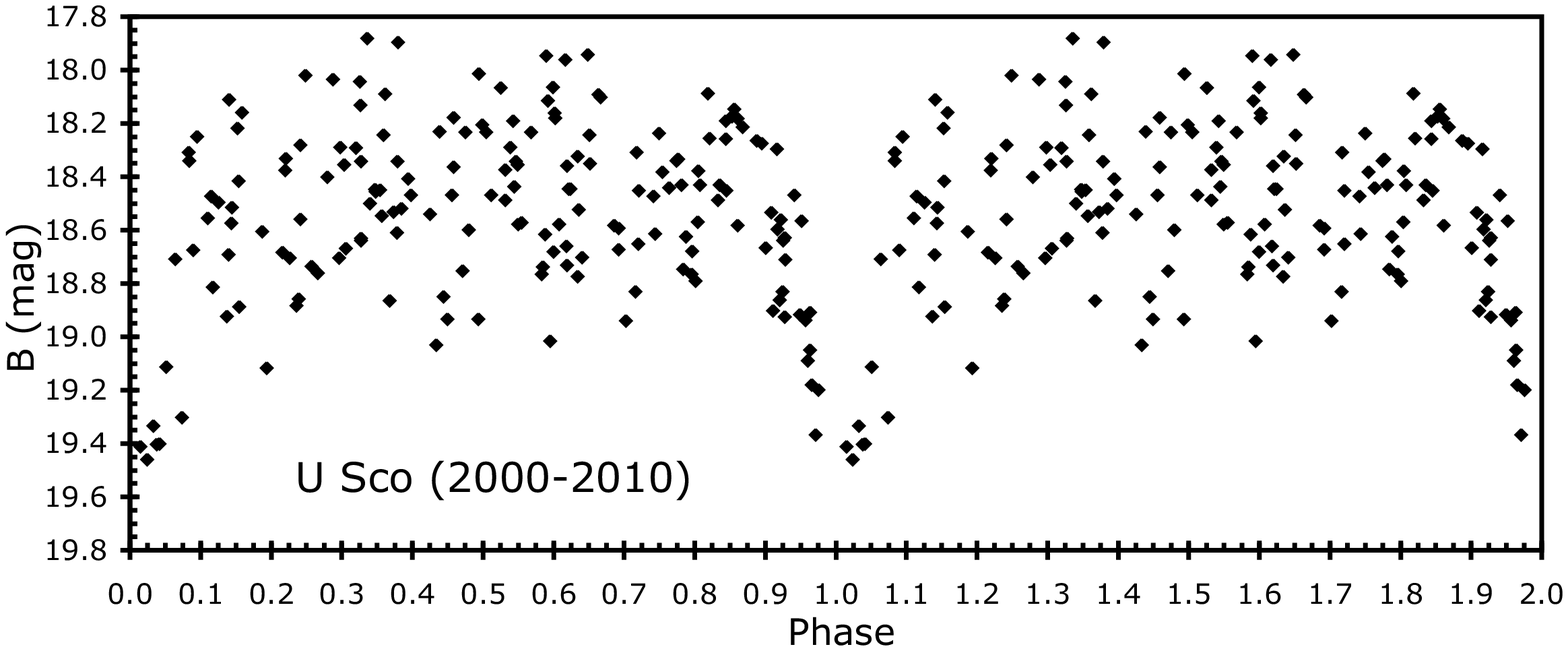}
\plotone{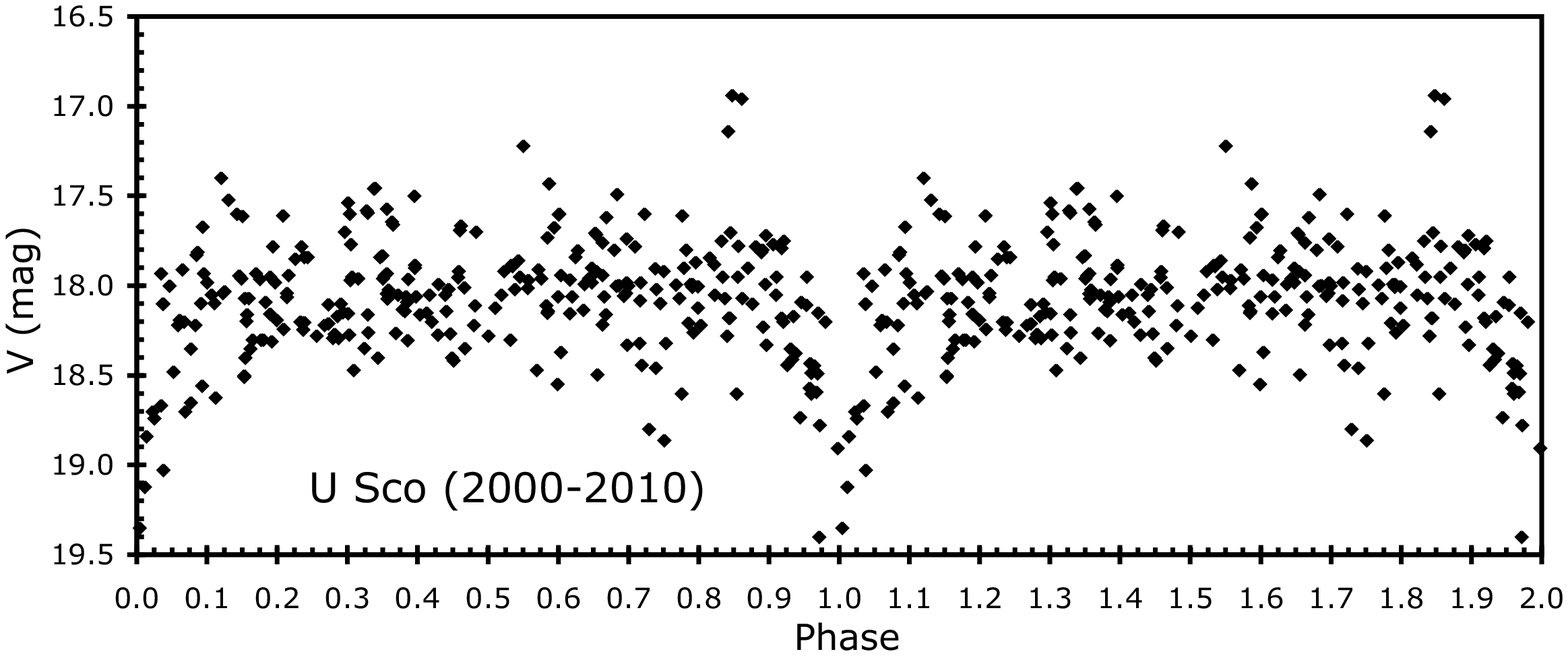}
\caption{
U Sco phased light curve in B and V.  The two panels show the B and V magnitudes as a function of U Sco's orbital phase for all observations from 2000 until the 2010 eruption.  Each magnitude is double plotted, once for phases 0-1 and a second time with unity added to the phase, so as to ease the visibility of the eclipse.  Both light curves show substantial short-term variations superposed on a flat light curve with eclipses.  The eclipses look rather ragged, the result of taking one isolated point from many eclipses over which the system is varying up and down.  (Time series through individual eclipses show a well-defined classic eclipse shape.)  No secondary eclipse is visible in either B or V.  Note that the scatter apparent outside of eclipse is much smaller during the eclipse, pointing to the flickering region being eclipsed.}
\end{figure}

\end{document}